\documentclass[prd,twocolumn,superscriptaddress,amsfonts,amssymb,amsmath,showpacs]{revtex4-2}
\usepackage{bm}
\usepackage{amsfonts}
\usepackage{latexsym}
\usepackage[latin1]{inputenc}
\usepackage{graphicx}
\usepackage{amsmath}
\usepackage{palatino}
\usepackage{mathpazo}
\usepackage{textcomp}
\usepackage{float}
\usepackage{booktabs}
\usepackage{dcolumn}
\usepackage{ragged2e}
\usepackage{hyperref}
\hypersetup{colorlinks,citecolor=blue,urlcolor=blue}
\usepackage{amsmath}
\usepackage{xcolor}
\usepackage{orcidlink}
\usepackage{epsfig}
\usepackage{cleveref}
\usepackage{caption}
\usepackage{subcaption}
\usepackage{commath}

\def\jnl@style{\it}
\def\aaref@jnl#1{{\jnl@style#1}}

\def\aaref@jnl#1{{\jnl@style#1}}

\def\aj{\aaref@jnl{AJ}}                   
\def\apj{\aaref@jnl{ApJ}}                 
\def\apjl{\aaref@jnl{ApJ}}                
\def\apjs{\aaref@jnl{ApJS}}               
\def\apss{\aaref@jnl{Ap\&SS}}             
\def\aap{\aaref@jnl{A\&A}}                
\def\aapr{\aaref@jnl{A\&A~Rev.}}          
\def\aaps{\aaref@jnl{A\&AS}}              
\def\mnras{\aaref@jnl{Mon.~Not.~Roy.~Astron.~Soc.}}             
\def\prd{\aaref@jnl{Phys.~Rev.~D}}        
\def\prc{\aaref@jnl{Phys.~Rev.~C}}  
\def\prl{\aaref@jnl{Phys.~Rev.~Lett.}}    
\def\qjras{\aaref@jnl{QJRAS}}             
\def\skytel{\aaref@jnl{S\&T}}             
\def\ssr{\aaref@jnl{Space~Sci.~Rev.}}     
\def\zap{\aaref@jnl{ZAp}}                 
\def\nat{\aaref@jnl{Nature}}              
\def\aplett{\aaref@jnl{Astrophys.~Lett.}} 
\def\apspr{\aaref@jnl{Astrophys.~Space~Phys.~Res.}} 
\def\physrep{\aaref@jnl{Phys.~Rep.}}      
\def\physscr{\aaref@jnl{Phys.~Scr}}       
\def\commat{\aaref@jnl{Comm.~Math.~Phys.}}              
\def\science{\aaref@jnl{Science}}               
\def\cqg{\aaref@jnl{Classical Quant.~Grav.}}            
\def\jpcs{\aaref@jnl{JPCS}}                                     
\def\ijmpd{\aaref@jnl{Int.~J.~Mod.~Phys.~D}}                    
\def\grg{\aaref@jnl{Gen.~Relat.~Gravit.}}               
\def\rpp{\aaref@jnl{Rep.~Prog.~Phys.}}          
\def\npa{\aaref@jnl{Nucl.~Phys.~A}}        
\def\lrr{\aaref@jnl{Living Rev.~Rel.}}                   
\def\jcap{\aaref@jnl{J.~Cosmology Astropart.~Phys.}}    
\def\rmp{\aaref@jnl{Rev.~Mod.~Phys.}}   
\def\epjc{\aaref@jnl{Eur.~Phys.~J.~C}} 
\def\plb{\aaref@jnl{~Phy.~Lett.~B}} 
\def\mpla{\aaref@jnl{Mod.~Phy.~Lett.~A}} 
\def\arxiv{\aaref@jnl{arxiv.org}}

\allowdisplaybreaks[1]

\begin{document}

\color{black}       

\title{Constraining Parameters for the Accelerating Universe in $f(R,\mathcal{L}_{m})$ Gravity}

\author{Y. Kalpana Devi\orcidlink{}}
\email{kalpanayengkhom123@gmail.com}
\affiliation{Department of Mathematics, Birla Institute of Technology and Science-Pilani,\\ Hyderabad Campus, Hyderabad-500078, India.}

\author{S.A. Narawade \orcidlink{0000-0002-8739-7412}}
\email{shubhamn2616@gmail.com}
\affiliation{Department of Mathematics, Birla Institute of Technology and Science-Pilani,\\ Hyderabad Campus, Hyderabad-500078, India.}

\author{B. Mishra\orcidlink{0000-0001-5527-3565}}
\email{bivu@hyderabad.bits-pilani.ac.in}
\affiliation{Department of Mathematics, Birla Institute of Technology and Science-Pilani,\\ Hyderabad Campus, Hyderabad-500078, India.}

\date{\today}

\begin{abstract}
\textbf{Abstract}: In the paper, we present an accelerating cosmological model in $f(R,\mathcal{L}_{m})$ gravity with the parameter constrained through the cosmological data sets. At the beginning, we have employed a functional form of $f(R,\mathcal{L}_{m}) =\frac{R}{2}+\alpha R^2+\mathcal{L}_{m}^\beta$, where $\alpha$ and $\beta$ are model parameters. This model is well motivated from the Starobinsky model in $f(R)$ gravity and the power law form of $f(\mathcal{L}_{m})$. The Hubble parameter has been derived with some algebraic manipulation and constrained by Hubble data and Pantheon$^{+}$ data. With the constraint parameters, present value of deceleration parameter has been obtained to as $q_{0}\approx-0.63$ with the transition at $z_{t}\approx0.7$. It shows the early deceleration and late time acceleration behaviour. The present value of other geometric parameters such as the jerk and snap parameter are obtained to be $j_{0}\approx0.78$ and $s_{0}\approx 0.1$ respectively. The state finder diagnostic test gives the quintessence behaviour at present and converging to $\Lambda$CDM at late times. Moreover the $Om(z)$ diagnostics gives negative slope which shows that the model favours the state finder diagnostic result. Also the current age of Universe has been obtained as, $t_{0} = 13.64~~Gyrs$. The equation of state parameter also shows the quintessence behaviour. Based on the present analysis, it indicates that the $f(R,\mathcal{L}_{m})$ gravitational theory  may be another alternative to study the dark energy models.
\end{abstract}

\maketitle

\section{Introduction} 
In gravitational interactions, General Relativity (GR) is the most commonly used theory. Levi-Civita has been used to describe gravity by using Riemannian geometry. Space time is measured by Ricci curvature $R$ and the geometry is nonmetricity-free and torsion-free. Astrophysical observations from supernovae of type Ia \cite{Bennett_2003, Spergel_2003, Spergel_2007}, cosmic microwave background anisotropies \cite{Perlmutter_1997, Perlmutter_1998, Perlmutter_1999, Riess_2004, Riess_2007}, large scale structure \cite{Hawkins_2003, Max_2004, Cole_2005}, baryon acoustic oscillations \cite{Eisenstein_2005}, and weak lensing \cite{Nojiri_2007} indicate that in the present epoch, the Universe is accelerating. The most promising feature of the Universe is the dominance of exotic energy component with large negative pressure, known as dark energy (DE). In general, the research devoted to modifying and extending GR has two main motivations: (i) the modified gravity hypothesis is based on cosmological grounds, and it is an effective explanation for the expansion of the Universe over time \cite{Capozziello_2011, Saridakis_2021} and; (ii) for purely theoretical reasons. In order to achieve a quantum gravity theory \cite{Stelle_1977}, it intends to enhance renormalizability of the GR. As described in \cite{Riess_1998, Perlmutter_1999, Spergel_2003, Eisenstein_2005, Betoule_2014, Ade_2015, Aghanim_2016} the observed accelerated expansion of the Universe has been modeled by the cosmological constant $\Lambda$. There have been numerous challenges to the $\Lambda$CDM (Cold Dark Matter) cosmological model. To explain the role of DE in cosmic acceleration, a number of alternative models have been proposed using general relativity (GR). One of the approach is, by replacing scalar curvature $R$ by an arbitrary function $R$ in Einstein-Hilbert action, referred as $f(R)$ gravity.

$f(R)$ theory of gravity one of the most viable and realistic theory among the various modifications of GR. It is considered as one of the most cosmologically important theory due to its higher order curvature invariant models. The late time cosmic expansion behaviour of the Universe can also be explained by $f(R)$ theory \cite{Carroll_2004}. Since it has been shown in \cite{Riess_1998} that $f(R)$ theory is equivalent to scalar-tensor theory, it is incompatible with solar system tests under weak fields. Nevertheless some models that also satisfies the solar system test are also discussed in \cite{Hu_2007, Sawicki_2007, Amendola_2008}. It is also discussed in Capozziello et al. \cite{Capozziello_2008a} that the matter dominating and dark energy dominating phases can be achieved from a $f(R)$ power low cosmological model. Various cosmological models shows the unification of early time inflation and late time cosmic acceleration \cite{Nojiri_2007a, Multamaki_2004, Multamaki_2007, Shamir_2010}. It can also describe the phase transition of the Universe from deceleration to acceleration. Energy conditions to locally homogeneous and isotropic Universe in $f(R)$ gravity have been discussed in \cite{Santos_2007}. For such certain reasons $f(R)$ theory has attract considerable interest.

$f(R,\mathcal{L}_{m})$ theory of gravity arises from an extension from $f(R)$ theory with an explicit coupling between matter Lagrangian density $(\mathcal{L}_{m})$ and the Ricci scalar ($R$). It was purposed in \cite{Bohmer_2007} and found out that the motion of a test particle was non-geodesic and an extra force arises orthogonal to the four-velocity due to the coupling. The physical implications of such force have been discussed under a generalized gravity model with an arbitrary coupling between the matter Lagrangian density and Ricci scalar  \cite{Harko_2008}. A more generalized study is done by considering the gravitational Lagrangian as a function of Ricci scalar $R$ and matter Lagrangian density $\mathcal{L}_{m}$ in \cite{Harko_2010b}. It is one of the most extended theory with the background of Riemann geometry. Various impacts of the non-minimal matter-geometry coupling on astrophysical and cosmological scenarios have been discussed in \cite{Harko_2010, Harko_2010a, Savvas_2009, Valerio_2007, Valerio_2009, Gonccalves_2023}. The non-geodesic equation of motion of the particles due to the additional force can lead to significant modification on the cosmological behaviour. The extra force has impact on the trajectory of the particles and it can potentially explain the accelerated expansion of the Universe without involving the dark matter term. It can also effect the expansion history of the Universe, alter the cosmic growth rate and structure formation. So it can be helpful to study such effects and understand the viability and implication of $f(R,\mathcal{L}_{m})$ gravity.\\

The article has been discussed in various sections as follows.: In Section \ref{secII} we have derived the equation of motion for spatially flat FLRW spacetime geometry in $f(R,\mathcal{L}_{m})$ gravity and introduced the $H(z)$ parameterization. Then, we have done the statistical analysis of the numerical solution for Hubble parameter with Monte Carlo Markov Chain (\textit{MCMC}) simulation and obtained the constrained free model parameters using observational datasets in Section \ref{secIII}. Using the constrained free model parameter values we studied the behaviour of cosmographic parameter in Section \ref{secIV}. In Section \ref{secV} we have performed various tests to check the validation of model containing the state finder diagnostics and the $Om(z)$ diagnostics. The results obtained and the conclusions are given in Sec.\ref{secVI}.

\section{Mathematical Formalism}\label{secII}

The action for $f(R,\mathcal{L}_m)$ gravity \cite{Harko_2010b},

\begin{equation}\label{cfrl1}
    S=\int d^4x \sqrt{-g}f(R,\mathcal{L}_m),
\end{equation}
where $g$, $R$ and $\mathcal{L}_m$ be respectively the metric determinant, Ricci scalar and matter Lagrangian; $8\pi G=1$. Applying the variational principle in Eq. \eqref{cfrl1}, one can obtain the field equations of $f(R,\mathcal{L}_m)$ gravity as,

\begin{multline}\label{Eq:field1}
    R_{\mu\nu}f_R+(g_{\mu\nu}\nabla_\mu\nabla^\mu-\nabla_\mu\nabla_\nu )f_R-\frac{f}{2}g_{\mu\nu} \\ 
    =\frac{1}{2}f_{\mathcal{L}_m}(T_{\mu\nu}-\mathcal{L}_mg_{\mu\nu}).
\end{multline}
For brevity, we represent $f\equiv f(R,\mathcal{L}_m)$, $f_R\equiv\partial f(R,\mathcal{L}_m)/\partial R$ and $f_{\mathcal{L}_m}\equiv\partial f(R,\mathcal{L}_m)/\partial\mathcal{L}_m$. We consider, $f(R,\mathcal{L}_{m})=f_{1}(R)+f_{2}(R)G(\mathcal{L}_{m})$, where $f_{1}(R)$ and $f_{2}(R)$ are arbitrary function of Ricci scalar and $G(\mathcal{L}_{m})$ is a function of matter Lagrangian density. When $f_{1}(R)=1$, $f_{2}(R)=1$ and $G(\mathcal{L}_{m})=\mathcal{L}_{m}$, Eq. \eqref{Eq:field1} reduces to the field equations of General Relativity(GR) whereas for $f_{2}(R)=1$ and $G(\mathcal{L}_{m})=\mathcal{L}_{m}$ it reduces to that of $f(R)$ gravity. Moreover, for linear coupling of matter and geometry, $G(\mathcal{L}_{m})=1+\lambda \mathcal{L}_{m}$, where $\lambda$ is a constant. Now, the contracting form of Eq. \eqref{Eq:field1} can be written as,
\begin{equation}\label{Eq:field2}
f_{R}R+3\nabla_\mu\nabla^\mu f_{R}-2f=f_{\mathcal{L}_{m}}\left(\frac{1}{2}T-2\mathcal{L}_{m}\right).
\end{equation}
From Eq. \eqref{Eq:field1} and Eq. \eqref{Eq:field2}, one can eliminate the term $\nabla_\mu\nabla^\mu f_{R}$ and the resulting equation becomes,
\begin{multline}
f_{R}\left(R_{\mu \nu}-\frac{1}{3}R g_{\mu \nu}\right)+\frac{f}{6}-\nabla_{\mu}\nabla_{\nu}f_{R} \\
=\frac{1}{2}f_{\mathcal{L}_{m}}\left(T_{\mu \nu}-\frac{1}{3}(T-\mathcal{L}_{m}) g_{\mu \nu} \right).
\end{multline}
Taking the covariant divergence of Eq. \eqref{Eq:field1}, with the use of the following mathematical identity
\begin{equation}
\nabla^{\mu}\left[R_{\mu \nu }f_{R}+( g_{\mu\nu}\nabla_\mu\nabla^\mu-\nabla_{\mu}\nabla_{\nu})f_{R}-\frac{f}{2}g_{\mu \nu}\right]=0,   \end{equation}
one can obtain the divergence of energy-momentum tensor $T_{\mu \nu}$ as,
\begin{eqnarray}
\nabla^{\mu}T_{\mu \nu}&=&\nabla^{\mu}\ln[f_{\mathcal{L}_{m}}](\mathcal{L}_{m}g_{\mu \nu}-T_{\mu \nu}), \nonumber \\
&=&2\nabla^{\mu}\ln[f_{\mathcal{L}_{m}}]\frac{\partial \mathcal{L}_{m}}{\partial g^{\mu \nu}}.
\end{eqnarray}
The requirement of the conservation of the energy-momentum tensor of matter ($\nabla^{\mu}T_{\mu \nu}=0$) provides an effective functional relation between the matter and Lagrangian density. Moreover, the conservation of the energy-momentum tensor yields

\begin{equation}\label{cfrl3}
	\nabla^\mu\ln f_{\mathcal{L}_{m}}=0.
\end{equation}

We consider the FLRW space-time as,
\begin{equation} \label{flrw}
ds^2=-dt^2+a^{2}(t)\left(dr^{2}+r^2 d\theta^2+r^2\sin^2\theta d\phi^2\right),
\end{equation} 
where $a(t)$ is the scaling factor with cosmic time $t$. The stress energy-momentum tensor for a perfect fluid can be written as,
\begin{align}\label{stressenergy}
    T_{\mu \nu}=(\textit{p} + \rho ) u_{ \mu } u_{ \nu }+ \textit{p} g_{ \mu \nu },
\end{align}
where $u_{\mu}$ is the four velocity vectors along the time directions, $\rho $ is the matter energy density and $\textit{p} $ is the isotropic pressure; $u_{\mu}$ satisfy the relations $u_{\mu} u_{\nu} g^{\mu \nu }=-1$. Using Eq. \eqref{flrw} and Eq. \eqref{stressenergy} in Eq. \eqref{Eq:field1}, the field equations of $f(R, \mathcal{L}_m)$ gravity can be obtained as,
\begin{eqnarray}
3H^2f_R+\frac{1}{2}(f-f_R R-f_{\mathcal{L}_m}\mathcal{L}_{m})+3H\dot{f_R}&=&\frac{1}{2}f_{\mathcal{L}_m}\rho, \nonumber\label{fieldequation1}\\ 
\dot{H}f_R+3H^2f_R-\ddot{f_R}-3H\dot{f_R}+\frac{1}{2}(f_{\mathcal{L}_m}\mathcal{L}_{m}-f)&=&\frac{1}{2}f_{\mathcal{L}_m}p,\nonumber \label{fieldequation2}\\
\end{eqnarray}
 where the Hubble parameter, $H=\frac{\dot{a}}{a}$. To solve the system [Eq. \eqref{fieldequation1}], we need to define a well motivated form of $f(R,\mathcal{L}_{m})$. Since $f(R,\mathcal{L}_{m}) \rightarrow f(R)+f(\mathcal{L}_{m})$, we consider $f(R)=R+\alpha R^2$ \cite{Starobinsky_1980} with a coefficient of $(1/2)$ in the $'R'$ term such that the resulting equations of motion correctly reduces to General Relativity under certain limits and $f(\mathcal{L}_{m})=\mathcal{L}_{m}^\beta$ inspired from power law model which shows similar background to $\Lambda CDM$ for small value and redshift and deviates slowly with increase in redshift \cite{Harko_2014}. Hence the form becomes
\begin{equation}\label{cosmological model}
f(R,\mathcal{L}_{m})=\frac{R}{2}+\alpha R^2+\mathcal{L}_{m}^\beta,
\end{equation}  
so that $f_R = \frac{\partial f(R,\mathcal{L}_{m})}{\partial R}=\frac{1}{2}+2\alpha R$, $f_{RR}=\frac{\partial f_R}{\partial R}=2\alpha$, $\dot {f_R}=\frac{\partial f_R}{\partial t}=2 \alpha \dot{R}$ and $f_{\mathcal{L}_{m}} =\beta \mathcal{L}_{m}^{\beta -1}$, where $\alpha$ and $\beta$ are constants and are to be constrained in order to get the accelerating behaviour. Considering $\mathcal{L}_m=\rho$ and the barotropic fluid $p=\omega \rho $ for a homogeneous and isotropic Universe and using Eq. \eqref{cosmological model}, the first Friedmann equation becomes,

\begin{equation} \label{firstfriedmann}
\frac{3}{2}H^2+108 \dot{H} H^2 \alpha +36 \Ddot{H} H^2 \alpha -18 \dot{H}^2 \alpha = (\beta- \frac{1}{2}) \rho ^\beta.
\end{equation}  
One can check that, Eq. \eqref{firstfriedmann} reduces to GR field equations when $\alpha = 0$ and $\beta = \frac{1}{2}$. Further we can express Eq. \eqref{firstfriedmann} as a function of redshift using the relation,

\begin{equation}
-dz=(1+z)H(z)dt,
\end{equation}
so that $\dot{H}=-(1+z)H(z)\frac{dH}{dz}$ and $\Ddot{H}=(1+z)^2 H^2(z)\frac{\partial^2 H}{\partial z^2}+(1+z)^2 H(z) (\frac{\partial H}{\partial z})^2 +(1+z)H^2(z)\frac{\partial H}{\partial z}$. Subsequently, Eq. \eqref{firstfriedmann} reduces to a second-order non-linear differential equations as,

\begin{multline}\label{firstfriedmannredshift}
H''+\frac{(2H-1)}{2H^2}(H')^2+\frac{(H-3)}{H(1+z)}H' \\
-\frac{\left(\beta-\frac{1}{2}\right) \left(3(H_0)^2(1+z)^3\Omega_{m0}\right)^\beta-\frac{3}{2}(H)^2}{24H^2(1+z)^2\alpha}=0.
\end{multline}
where prime (') denotes the derivative in redshift, $H_0$ be the present value of the Hubble parameter and $\Omega_{m0}$ denotes the present density parameter for matter dominated phase and the radiation era case is neglected in the further analysis. Now, we shall solve Eq. \eqref{firstfriedmannredshift} numerically to find the Hubble parameter so that the dynamical parameters and other geometrical parameters of the models can be analysed.

\section{The Cosmological Datasets and The Constrained Parameters}\label{secIII}
\begin{figure}[htbp]
    \centering
    \begin{subfigure}{0.45\textwidth}
        \centering
        \includegraphics[width=\textwidth]{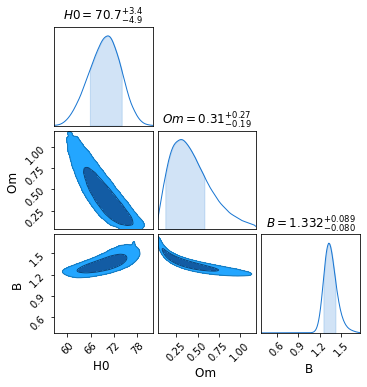}
        \caption{Using Hubble datasets}
        \label{fig:hzplot}
    \end{subfigure}
    \hfill
    \begin{subfigure}{0.45\textwidth}
        \centering
        \includegraphics[width=\textwidth]{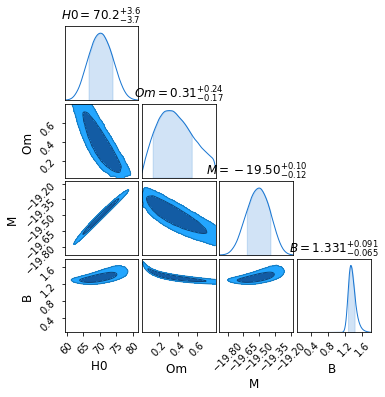}
        \caption{Using Hubble+Pantheon$^{+}$ datasets}
        \label{fig:hp1}
    \end{subfigure}
    \caption{Corner plots of constrained model parameters using (a) Hubble datasets on {\bf upper panel} (b) combined Hubble and Pantheon$^{+}$ datasets on {\bf lower panel} }
    \label{fig1}
\end{figure}

In this study we use Hubble and Hubble+Pantheon$^{+}$ datasets to constrained the model parameters using Markov Chain Monte Carlo \textit{(MCMC)} simulation for parameter estimation and exploration of the posterior distribution of Bayesian models. 
\begin{itemize}
    \item \textbf{\textit{Hubble Dataset :}} The observational value of the Hubble parameters give information about the expansion of the Universe. For finding our Hubble parameter we are using thirty-one data points which is obtained by the cosmic chronometers (CC), also known as the differential age technique, which is independent of the fiducial model. This technique was proposed by Jimenez and Loeb in \cite{Jimenez_2002}. The basic idea of this technique involve the measurement of age difference between two passively evolving early type galaxies by assuring that they are formed at the same time but with small difference at the redshift interval i.e. $\left(\frac{dz}{dt}\right)$. It provides us information about the Hubble function at various redshifts. We use thirty-one data points from redshift range between $0.07 < z < 1.965$. The corresponding $\chi^2_H$ is given by 
    \begin{equation}
       \chi^2_H(\Theta)= \sum_{i=1}^{31} \frac{((H(z_i,\Theta)-H_{obs}(z_i))^2}{\sigma^2_H(z_i)}.
    \end{equation}
    Here $H(z_i,\Theta)$ is the theoritical value of Hubble parameter which we obtain from solving equation \eqref{firstfriedmannredshift} and $\Theta$ represents the model parameters, $H_{obs}(z_i)$ is the observed value of Hubble parameter of the corresponding redshift and $\sigma_H(z_i)$ is the observational error.

    The Hubble dataset \cite{Narawade_2023} is given below which is used for the analysis of the behaviour of the numerical solution of Eq. \eqref{firstfriedmannredshift}

   \item \textbf{\textit{Pantheon$^{+}$ Data :}} The dataset consists of compilation of $1701$ light curves of $1550$ distinct Type Ia supernovae (SNe Ia) with relative luminosity distance ranging from $z= 0.00122$ to $2.2613$ \cite{Brout_2022}. The magnitudes of parameters such as (stretch of the light curve, the colour at the maximum brightness and the steller mass of the host galaxy) provided data is free from the systematic errors. We can write the the values of the distance moduli as, 
   \begin{equation}\label{distancemod}
       \mu (z_i,\Theta) = 5 log_10[D_L(z_i,\Theta]+M,
   \end{equation}
where $M$ is the absolute luminosity distance of a known start also act as a nuisance parameter in our \textit{MCMC} simulation, it is added since the apparent magnitude of each luminosity distance needs to be calibrated. The corresponding luminosity distance $D_L$ at each corresponding redshift $z_i$ can be written as,
\begin{equation}\label{luminosity}
    D_L(z_i,\Theta)=c(1+z_i)\int_ {0}^{z_i} \frac{dz'}{H(z',\Theta)},
\end{equation} 
where $H$ is the Hubble parameter which we will be estimating using \textit{MCMC} simulation. The corresponding $\chi^2_{SN} $ can be written as,
\begin{equation}
    \chi^2_{SN}(\Theta)=(\Delta\mu(z_i,\Theta))^T C^{-1}_{SN}\Delta\mu(z_i,\Theta)+ln\left(\frac{S}{2\pi}\right)-\frac{k^2(\Theta)}{S},
\end{equation}
where $C_{SN}$ is the total covariance matrix, $S$ is the sum of all the components of $C^{-1}_{SN}$ and $$k(\Theta) =\big(\mu(z_i,\Theta)-\mu_{obs}(z_i)\big)^TC^{-1}_{SN}.$$
\end{itemize}
Markov Chain Monte Carlo \textit{(MCMC)} analysis is used to constrain the model parameters using the above mentioned datasets and $\chi^2$ minimization technique. The constrained values will be useful in interpreting various cosmological behaviour. We have assumed the value of $\alpha=0.01$. \textit{MCMC} is used to obtain the posterior distribution of parameters from a complex probability distributions using the observational data and a prior distribution idea. FIG.- \ref{fig1} corner plot shows the joint and marginal distributions of the model parameters. The diagonal elements of the plots shows the marginal distribution of each model parameters, the sharpness and width of each plot represents the well constrained and uncertainty in constraining the parameter values. The off diagonal plots shows the joint distribution of the parameters, elliptical contours of this plots shows the correlation between the parameters. The blue and light blue region of the joint distribution plots shows the 2- $\sigma$ and 1-$\sigma$ confidence level. FIG.- \ref{fig:hzplot} and FIG.- \ref{fig:hp1} shows the constrained corner plot of the model parameters using Hubble data and Hubble+Pantheon$^{+}$ data respectively. The constrained parameter values are given in TABLE - \ref{Table:2}.

\begin{table}[h!]
\centering
\caption{Constrained cosmological parameter values from different datasets}\label{Table:2}
\begin{tabular}{l c c}
\toprule
\textbf{Parameter} & \textbf{Hubble Data} & \textbf{Hubble + Pantheon$^{+}$ Data} \\
\midrule
$H_0$ & $70.7^{+3.4}_{-4.9}$ & $70.2^{+3.6}_{-3.7}$ \\
\\
$\Omega_{m0}$ & $0.31^{+0.27}_{-0.19}$ & $0.31^{+0.24}_{-0.17}$ \\
\\
$M$ & - & $-19.50^{+0.10}_{-0.12}$ \\
\\
$\beta$ & $1.332^{+0.089}_{-0.080}$ & $1.331^{+0.091}_{-0.065}$ \\
\bottomrule
\end{tabular}
\end{table}

\begin{figure}[H]
    \centering
    \includegraphics[width=9cm,height=8cm]{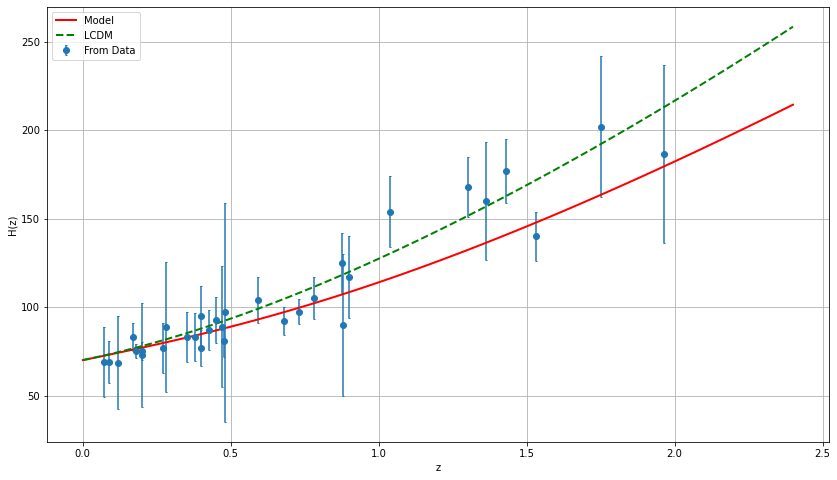}
    \caption{$H(z)$ vs redshift plot with errorbars from the Hubble data points in TABLE- \ref{Table1}. The solid red line is for the model and the dotted green line for the $\Lambda$CDM with errorbar plots for the $32$ Hubble data points}
    \label{fig:2}
\end{figure}

\begin{figure}[H]
    \centering
    \includegraphics[width=9cm,height=8cm]{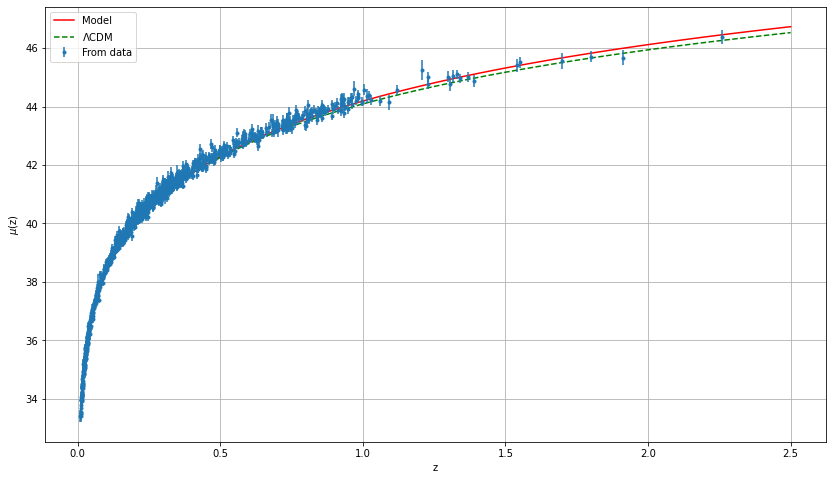}
    \caption{$\mu(z)$ vs redshift with error bar plot using Pantheon$^{+}$ data. The red line is for the model and the green dotted is the plot for $\Lambda$CDM.}
    \label{fig:3}
\end{figure}

Using the constrained model parameter values form TABLE - \ref{Table:2} we can check the viability of the cosmological model mentioned in Eq. \eqref{cosmological model} in comparison with the standard $\Lambda$CDM model. In FIG.- \ref{fig:2} we have plotted errorbar plot of Hubble parameter $H(z)$ with redshift (z) using Hubble data points from TABLE- \ref{Table1} . The dotted green lines represent the $\Lambda$CDM and solid red line is for the model with constrained parameter values from TABLE- \ref{Table:2}. The plot shows deviation of the model behaviour form $\Lambda$CDM with increase in redshift value. We calculated the value of distance modulus form Hubble parameter value using equation {\eqref{distancemod}} and {\eqref{luminosity}} and see the variation of distance modulus with redshift. FIG.- \ref{fig:3} is the plot for distance modulus $\mu(z)$ vs redshift $(z)$ with errorbar plot using the Pantheon$^{+}$ dataset \cite{Brout_2022}. The solid red line which passes through the error bars is for the assumed model with constrained model parameter values and dotted green lines is for the $\Lambda$CDM model. In both FIG.- \ref{fig:2} and FIG.- \ref{fig:3}
we obtained the plots by considering the model parameter values as $H_0= 70$, $\Omega_{m0}= 0.3$ and $\beta=1.3$. In FIG.- \ref{fig:3},  the error bars are plotted with corrected value of $M=+19.5$ both the plots shows a compatible behaviour of the model with $\Lambda $CDM with slight deviation as the value of redshift increases.

\section{Cosmographic Parameters}\label{secIV}
According to the cosmological principle, the scale factor seems to be the only degree of freedom governing the Universe. We may expand the scale factor $a(t)$ in a Taylor series around $a_0$ as \cite{Aviles_2012, Gruber_2014, Weinberg_1971, Visser_2004, Visser_2005},
\begin{eqnarray}
a(t)=a_0+\sum_{n=0}^{\infty} \frac{1}{n!}\frac{d^na}{dt^n}\mid_{(t=t_0)} (t-t_0)^n,\label{eq:32}
\end{eqnarray}
where $t_0$ is the present cosmic time and $n=1,2,3, \cdots$ is an integer. The coefficients of expansion are referred to as the cosmographic coefficients. In fact, the cosmographic coefficients contain different order of derivatives of the scale factor and therefore may provide a better geometrical description of the model. At any time $t$, we may define some geometrical parameters as,
\begin{equation*}
H = \frac{\dot{a}}{a}, \quad\quad
q = -\frac{a\ddot{a}}{\dot{a}^2}, \quad\quad
j = \frac{\dddot{a}}{aH^3}, \quad\quad
s = \frac{a^{(4)}}{aH^4},
\end{equation*}
where $a^{(4)}$ is the fourth order derivatives of the scale factor. An integral part of the evolution of the Universe is the Hubble parameter $H$, which indicates how rapidly it is expanding. It is necessary to have a positive Hubble parameter if the Universe expands. The negative and positive signs of the deceleration parameter respectively provide information about the acceleration and deceleration of the Universe. These parameters can be expressed in redshift by using \eqref{firstfriedmannredshift} as,
 \begin{eqnarray}\label{cosmicparameter}
 q(z) &=& -1+\frac{(1+z)H'(z)}{H(z)}, \nonumber\\
 j(z) &=& 2q(z)^{2}+q(z) + (1+z)q'(z), \nonumber\\
 s(z) &=& \frac{j(z)-1}{3\left(q(z)-\frac{1}{2}\right)}, \quad q\neq \frac{1}{2}.
 \end{eqnarray}
\begin{figure}[H]
    \centering
    \includegraphics[width=8cm,height=8cm]{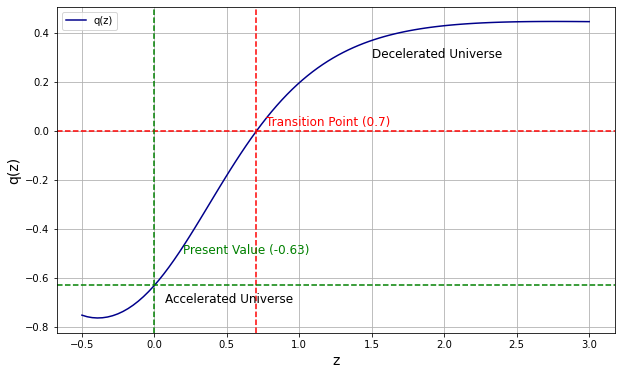}
    \caption{Behaviour of deceleration parameter in redshift.}
    \label{decelerationparameter}
\end{figure}
 
The value of $q$ (with $q_{0}$ representing the present value) dictates the behavior of the Universe as: a decelerating, expanding Universe occurs when $q_{0}>0$, characteristic of a pressure-less barotropic fluid or matter-dominated Universe, likely representing the early Universe, though current observations do not support a positive $q_{0}$. The current state of the Universe is an expanding and accelerating one, corresponding to $-1<q_{0}<0$. A value of $q_{0}=-1$ indicates a Universe dominated by a de Sitter fluid, relevant to the inflationary period of the very early Universe. From Eq. \eqref{cosmicparameter}, the present value of the jerk parameter $j_{0}$, is given by $j_{0} = 2q_{0}^{2}+q_{0}+ q'_{0}$. We consider $-1<q_{0}<-0.5$, which requires $2q_{0}^{2}+q_{0}>0$. Thus, if $q_{0}<-0.5$, $j_{0}$ depends on the sign of the variation of $q$. In other words, when $j_{0}$ is negative, the Universe remains in its present accelerated phase without any change, suggesting that dark energy has consistently influenced the dynamics since the dawn of time. Acceleration parameter stabilizes smoothly at a precise value when $j_{0}$ is zero. When $j_{0}$ is positive, there was a distinct point in the evolution of the Universe when acceleration began, marking a transition redshift where influence of dark energy became significant. The expansion of the Universe slope changes with the sign of $j_{0}$, which suggests additional cosmological factors. Measuring the transition redshift $z_{tr}$ directly is crucial to constraining the dark energy equation of state. 

\begin{figure}[H]
    \centering
    \includegraphics[width=8cm,height=8cm]{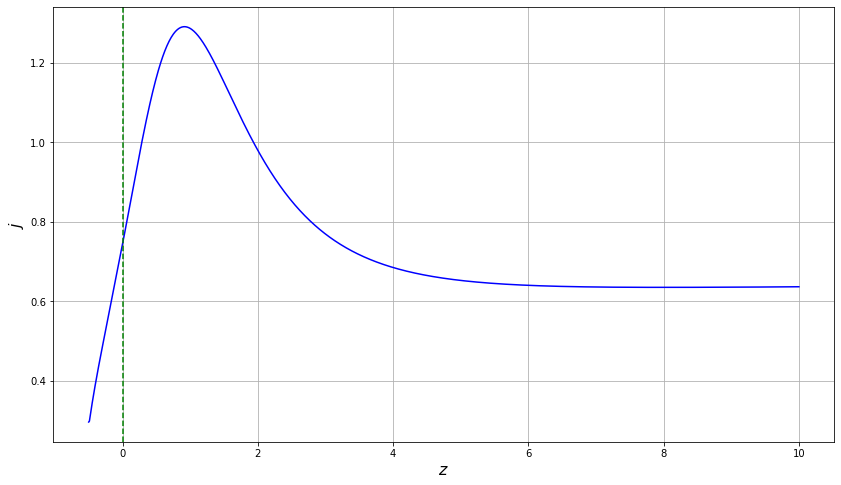}
    \caption{Behaviour of jerk parameter in redshift.}\label{jerkparameter}
\end{figure}
\begin{figure}[H]
    \centering
    \includegraphics[width=8cm,height=8cm]{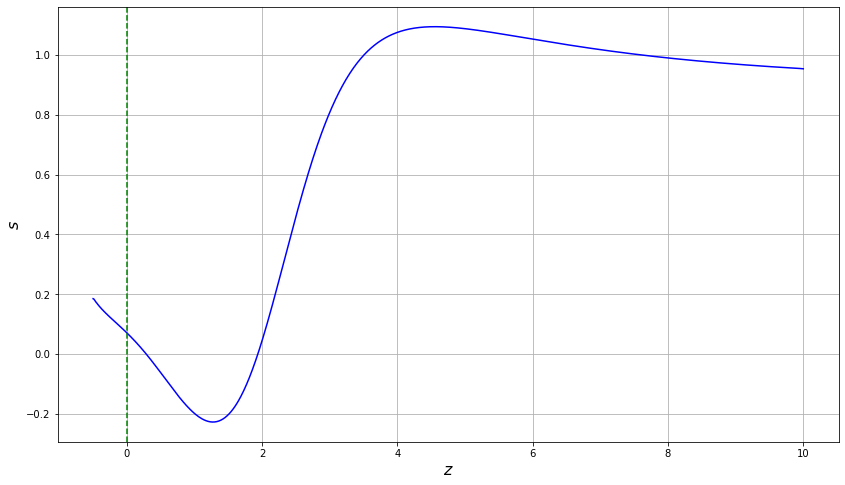}
    \caption{Behaviour of snap parameter in redshift.}\label{snapparameter}
\end{figure}

It is necessary to study the dynamics of deceleration parameter in order to understand the expansion nature of the Universe. In FIG.- \ref{decelerationparameter} we have plotted the change in deceleration parameter with respect to redshift using the constrained parameter values in TABLE- \ref{Table:2} from Hubble and Hubble+Pantheon$^{+}$ dataset as they give almost the same constrained parameter values. As mentioned above for a expanding Universe, Hubble parameter has to be positive and from the formula of $q(z)$ in Eq. \eqref{cosmicparameter} we can say that $q(z)$ has to be negative. In the FIG.- \ref{decelerationparameter}, $z=0$ indicates the present time of the Universe and the value of $q(z)$ is $q_{0}=-0.63$ shows the accelerating phase of the Universe.  The recently performed measurements have determined that the value of the deceleration parameter for the current cosmic epoch is within the range of $q_{0} = -0.528_{-0.088}^{+0.092}$ \cite{Gruber_2014} and transition from deceleration to acceleration at $z_{t} = 0.60_{-0.12}^{+0.21}$ \cite{Yang_2020, Capozziello_2015}. As we can see from FIG.- \ref{decelerationparameter}, the deceleration parameter shows early deceleration to late time acceleration, where as the transition happens at $z_{t} = 0.70$. In FIG.- \ref{jerkparameter}, we have plotted the jerk parameter$(j)$ with respect to redshift, jerk parameter is a dimensionless parameter obtained form the coefficient fourth term in Eq. \eqref{eq:32}. From Fig.- \ref{jerkparameter} we can see the dynamical nature of the model showing constant rate of acceleration expansion of the Universe at early time. At late times, time varying acceleration with $(j>0)$ indicating increasing acceleration expansion, similar to the behavior of $\Lambda CDM$ model. We have also studied the $(s-z)$ plot in FIG.- \ref{snapparameter} to understand more about the model description of dynamics of dark energy.  We can observe that $s>0$ at early time shows the quintessence phase of the Universe and the trajectory of the plot converges to $\Lambda CDM$ in late time with $s=0$ which again explain the accelerating expansion behaviour of the Universe.

\section{Test for the validation}\label{secV}
The validity of any cosmological model can be checked through some theoretical and observational tests. In this discussion, we will examine several cosmological tests that may be employed to authenticate our derived model. Also calculated the age of the Universe to validate the model.

\noindent\textbf{\textit{EoS Parameter :}} Describes the relationship between cosmic fluid pressure and energy density using the equation of state (EoS) parameter. Generally represented as $\omega$, this parameter governs the expansion of the Universe and provides insight into its evolution stages. The value of $\omega$ may deviate from the standard cosmological constant case ($\omega=-1$), indicating exotic behaviors such as phantom energy ($\omega < -1$) or quintessence ($-1 <\omega < -1/3$). Understanding the EoS parameter in this gravity is essential for predicting the fate of the Universe whether it will experience accelerated expansion or deceleration in current cosmological research and observational studies.
\begin{figure}[H]
    \centering
    \includegraphics[width=8.5cm,height=8cm]{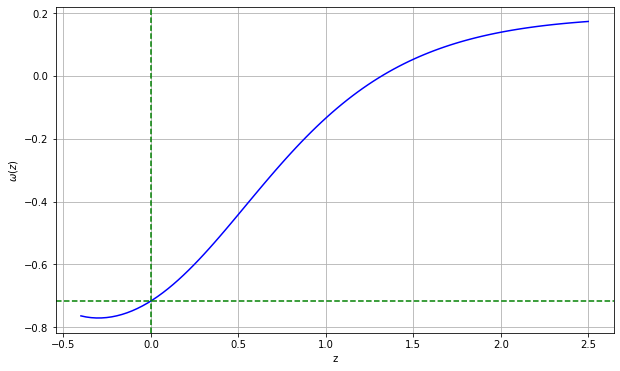}
    \caption{Behaviour of EoS parameter in redshift.}
    \label{eos}
\end{figure}
Using the Eq. \eqref{fieldequation2} and Eq. \eqref{firstfriedmannredshift}, the EoS parameter can be expressed. FIG.- \ref{eos} illustrates the dynamic nature of $\omega$ that can change significantly from an early to a late epoch. With the value of $\omega = -0.72$ at $z=0$, the EoS parameter shows the quintessence behavior at present and converges to $\Lambda$CDM in late time.\\

\noindent\textbf{\textit{Statefinder Diagnostic :}} The state finder pair $(j,s)$ characterizes the properties of dark energy in the model independent manner. Sahni et al \cite{Sahni_2003, Alam_2003} have proposed this diagnostic with the classification: (i) $(j=1,~s=0)$ corresponds to the $\Lambda$CDM model, (ii) $(j<1,~s>0)$ indicates Quintessence, (iii) $(j>1,~s<0)$ represents Chaplygin Gas, and (iv) $(j=1,~s=1)$ signifies SCDM. So, the state finder pair helps to distinguish different dark energy models. 
\begin{figure}[H]
    \centering
    \includegraphics[width=8cm,height=8cm]{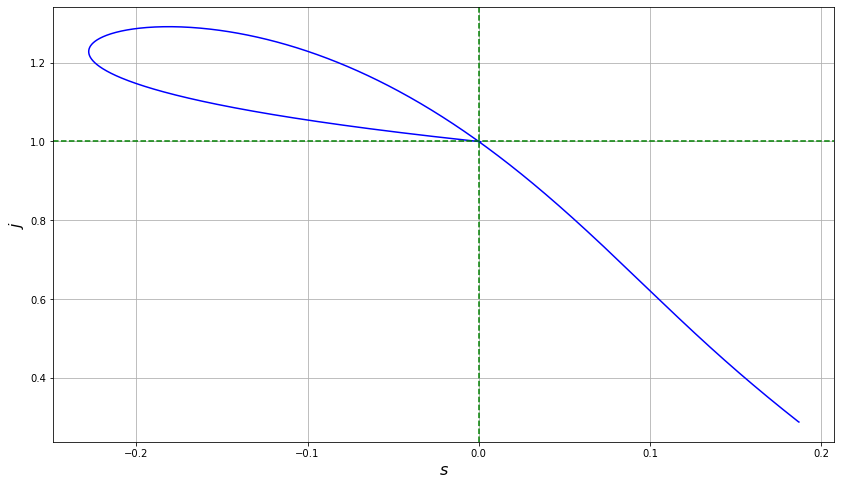}
    \caption{Behaviour of state finder pair in redshift.}
    \label{statefinder}
\end{figure}
\noindent One can see the behaviour of the state finder pair in FIG.- \ref{statefinder}, which is used to determine different dark energy model of the expansion Universe. In this method $j$ and $s$ are the two diagnostic parameters, the $j$-parameter gives us the rate of change in acceleration or deceleration of the Universe while the $s$-parameter helps in distinguishing different dark energy model. From this plot we can see the dynamics of the model with the constrained parameter values from the dataset, in early time the model shows Quintessence behaviour with $(j<1)$ for $(s>0)$ and the $j-s$ pair converges to $\Lambda$CDM in the late time at the fix point where $(s=0)$ and $(j=1)$ which shows the late time accelerating expanding behaviour of the Universe.

\noindent\textbf{\textit{$Om(z)$ Diagnostic :}} The $Om(z)$ diagnostic has been introduced as an alternative approach to test the accelerated expansion of the Universe with the phenomenological assumption, EoS, $p=\rho\omega$ filling the Universe with the perfect fluid. This diagnostic tool distinguishes the standard $\Lambda$CDM model from other dark energy models such as quintessence and phantom. Also, there are evidences available in the literature on its sensitiveness with the EoS parameter \cite{Ding_2015, Zheng_2016, Qi_2018}. The nature of $Om(z)$ slope differs between dark energy models because: the positive slope indicates the phantom phase $\omega < -1$, and the negative slope indicates the quintessence region $\omega > -1$. The $Om(z)$ diagnostic can be defined as, 
\begin{equation*}
Om(z) = \frac{E^{2}(z) - 1}{(1+z)^{3}-1} , \hspace{0.5cm}  E(z)= \frac{H(z)}{H0}
\end{equation*}
\begin{figure}[H]
    \centering
    \includegraphics[width=8cm,height=8cm]{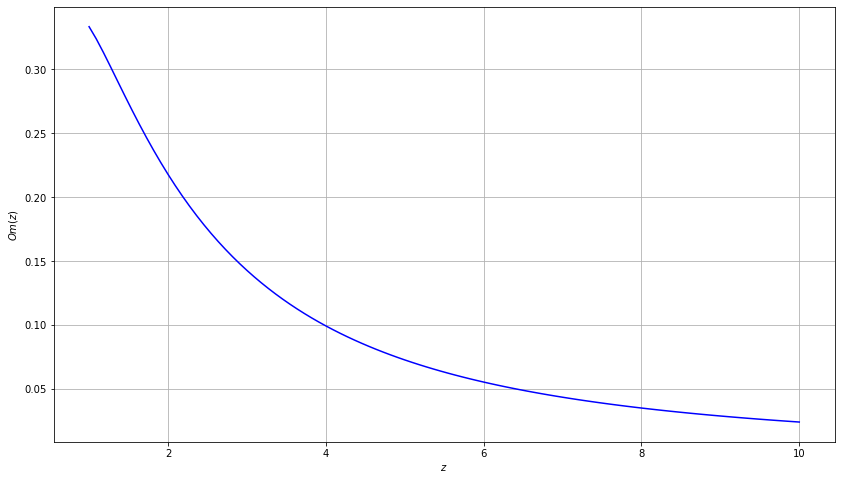}
    \caption{Behaviour of $Om(z)$ in redshift.}
    \label{Om}
\end{figure}
 In FIG.- \ref{Om}, the growth of $Om(z)$ has been illustrated which shows that at later stages, it  tends to support the viability of decaying
 dark energy models (quintessence dark energy at late time). 
 
\noindent\textbf{\textit{Age of the Universe :}} We can also calculate the age of the Universe using Hubble time as discussed in \cite{VAGNOZZI202227}. The age of the Universe as any value of redshift $(z)$ is given by 

\begin{equation*}
    t_U(z) = \int_{z}^{\infty} \frac{d\tilde{z}}{(1 + \tilde{z}) H(\tilde{z})}.
\end{equation*}
The present age of the Universe is given by the reciprocal of the Hubble constant $H_0= H(z=0)$. We can compute the value of age of Universe using the numerically calculated value Hubble parameter from \eqref{firstfriedmann} and the below formula
\begin{equation}
\begin{aligned}
H_0 t_0 &= \lim_{z \to \infty} \int_{0}^{z} \frac{d\tilde{z}}{(1 + \tilde{z}) E(\tilde{z})},\nonumber\\
t_0 &= \left[ \int_{0}^{z} \frac{d\tilde{z}}{(1 + \tilde{z}) E(\tilde{z})} \right] \times \left( \frac{1}{H_0}\right)
\end{aligned}
\end{equation}

\begin{figure}[H]
    \centering
    \includegraphics[width=8cm,height=8cm]{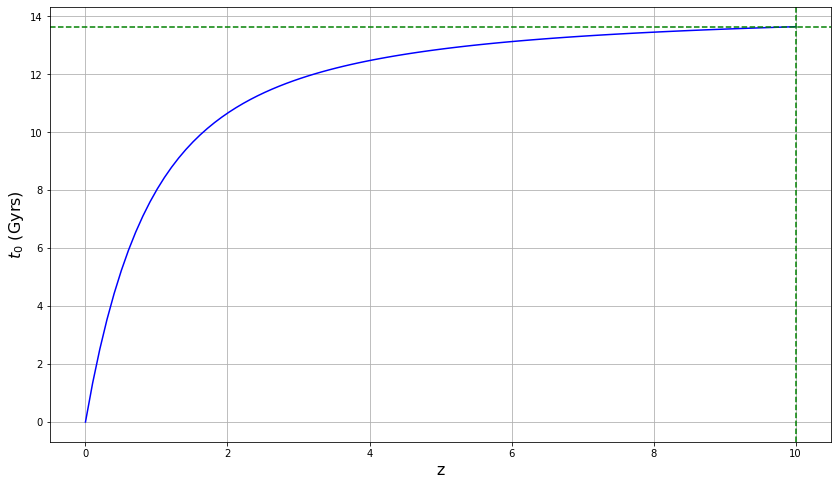}
    \caption{Hubble time $t_U(z)$ in redshift in terms of years.}
    \label{age}
\end{figure}
In FIG.- \ref{age} we have plotted age of the Universe in terms of years with respect to redshift. At a very large value of redshift the age of the Universe calculated from Hubble time is converging in $t_0 \approx 13.64~~ Gyrs$, quite close to the age calculated from the Planck result $t_{0} = 13.78 \pm 0.020~~Gyrs$. So, the results obtained in the model are in consistent with the present prescribed value according to cosmological observations.

\section{Conclusion}\label{secVI}
In this work, we have considered a functional form of the type $f(R)+f(\mathcal{L}_{m})$ in the $f(R,\mathcal{L}_{m})$ gravity, which is extension of the $f(R)$ gravity. The functional form of the $f(R,\mathcal{L}_{m})$ is inspired from the Starobinsky model in $f(R)$ gravity theory and $f(\mathcal{L}_{m})$ from powerlaw model and studied the dynamics of Universe using $f(R,\mathcal{L}_{m})$ gravity theory. After calculating the field equation for $f(R,\mathcal{L}_{m})$ using the gravitional action, the parametrized $H(z)$ is obtained using the field equations and the flat FLRW metric space time. Due to complexity of the differential equation for $H(z)$, the solution is obtained numerically for the $H(z)$.

The \textit{MCMC} analysis is used to estimate the free model parameters through the $32$ Hubble data points in the range $0.07 < z < 1.965$ and the $1701$ Pantheon$^{+}$ data points (light curves) in the range $0.00122 < z < 2.2613$ and the $\chi^2$ minimization technique. The constrained parameter values are shown in Table-\ref{Table:2} with the $1-\sigma$ confidence error. 
The comparison between the numerical solution for $H(z)$ and the $\Lambda$CDM model using error bar plots for Hubble data set and Pantheon$^{+}$ data set has been done. Moreover the geometrical parameters have been calculated. The present value for the deceleration parameter is obtained as $q_{0}\approx -0.63$, where the transition from deceleration to acceleration happens at the point $z_{t} \approx 0.70$. Whereas the present values for jerk parameter [$j_{0} = $] and snap parameter [$s_{0} = $] shows the quintessence behaviour of the model at present epoch.

We have calculated the EoS parameter for the model, which gives present value $\omega_{0}=-0.72$ showing the quintessence phase of the Universe. Also, we performed state finder diagnostic and $Om(z)$ diagnostic validation tests to the viability of the model. We obtained that the dynamics of the model shifting from early time quintessence phase and converging to late time $\Lambda$CDM model using state finder diagnostics. Moreover the $Om(z)$ diagnostics favors to the quintessence behaviour of the model. Finally we have calculated the age of the Universe using Hubble time and according to the model we have assumed the present age of the Universe is coming out to be $t_0\approx 13.64~~ Gyrs$ which is close to the calculated value from the Planck i.e $t_0=13.78\pm 0.020~~Gyrs$. So, overall this model shows a viable behaviour and consistent value with the observed data sets.

\section*{Acknowledgement} BM acknowledges the support of Council of Scientific and Industrial Research (CSIR) for the project grant (No. 03/1493/23/EMR II).\newpage

\section*{Appendix}
\begin{widetext}

    \begin{table}[H]
    \centering
    \caption{The observational Hubble dataset.}
    \resizebox{\textwidth}{!}{%
    \begin{tabular}{cccc||cccc}
    \toprule
    $\bm{Sr. No.}$ & $\bm{Redshift~(z)}$ & $\bm{H(z)~[km/s/Mpc]}$ & $\bm{\sigma_{H}~[km/s/Mpc]}$  & $\bm{Sr. No.}$ & $\bm{Redshift~(z)}$ & $\bm{H(z)~[km/s/Mpc]}$ & $\bm{\sigma_{H}~[km/s/Mpc]}$ \\
    \midrule
    1 & 0.07 & 69.0 & 19.6  & 17 & 0.4783 & 80.9 & 9.0  \\
    2 & 0.09 & 69.0 & 12.0  & 18 & 0.48 & 97.0 & 62.0  \\
    3 & 0.12 & 68.6 & 26.2 & 19 & 0.593 & 104.0 & 13.0 \\
    4 & 0.17 & 83.0 & 8.0 & 20 & 0.68 & 92.0 & 8.0 \\
    5 & 0.179 & 75.0 & 4.0  & 21 & 0.75 & 98.8 & 33.6  \\
    6 & 0.199 & 75.0 & 5.0  & 22 & 0.781 & 105.0 & 12.0  \\
    7 & 0.20 & 72.9 & 29.6  & 23 & 0.875 & 125.0 & 17.0  \\
    8 & 0.27 & 77.0 & 14.0  & 24 & 0.88 & 90.0 & 40.0  \\
    9 & 0.28 & 88.8 & 36.6  & 25 & 0.9 & 117.0 & 23.0  \\
    10 & 0.352 & 83.0 & 14.0  & 26 & 1.037 & 154.0 & 20.0  \\
    11 & 0.38 & 83.0 & 13.5  & 27 & 1.3 & 168.0 & 17.0  \\
    12 & 0.4 & 95.0 & 17.0 & 28 & 1.363 & 160.0 & 14.0 \\
    13 & 0.4004 & 77.0 & 10.2  & 29 & 1.43 & 177.0 & 40.0  \\
    14 & 0.425 & 87.1 & 11.2 & 30 & 1.53 & 140.0 & 18.0  \\
    15 & 0.445 & 92.8 & 12.9  & 31 & 1.75 & 202.0 & 14.0  \\
    16 & 0.47 & 89.0 & 49.6  & 32 & 1.965 & 186.5 & 50.4  \\
       
    \bottomrule
    \end{tabular}
    }
    \label{Table1}
    \end{table}
\end{widetext}

\section*{References}
\bibliographystyle{JCAP.bst}
\bibliography{Reference}

\providecommand{\href}[2]{#2}\begingroup\raggedright\begin{thebibliography}{10}

\bibitem{Bennett_2003}
C.L.~Bennett, M.~Halpern, G.~Hinshaw, N.~Jarosik et~al., \emph{{First-Year
  Wilkinson Microwave Anisotropy Probe (WMAP)* Observations: Preliminary Maps
  and Basic Results}}, \href{https://doi.org/10.1086/377253}{\emph{Astrophys.
  J. Supp. Ser.} {\bfseries 148} (2003) 1}.

\bibitem{Spergel_2003}
D.N.~Spergel, L.~Verde, H.V.~Peiris, E.~Komatsu et~al., \emph{{First-Year
  Wilkinson Microwave Anisotropy Probe (WMAP)* Observations: Determination of
  Cosmological Parameters}},
  \href{https://doi.org/10.1086/377226}{\emph{Astrophys. J. Supp. Ser.}
  {\bfseries 148} (2003) 175}.

\bibitem{Spergel_2007}
D.N.~Spergel, R.~Bean, O.~Dor\'e, M.R.~Nolta et~al., \emph{{Three-Year
  Wilkinson Microwave Anisotropy Probe (WMAP) Observations: Implications for
  Cosmology}}, \href{https://doi.org/10.1086/513700}{\emph{Astrophys. J. Supp.
  Ser.} {\bfseries 170} (2007) 377}.

\bibitem{Perlmutter_1997}
S.~Perlmutter, S.~Gabi, G.~Goldhaber, A.~Goobar et~al., \emph{{Measurements* of
  the Cosmological Parameters $\Omega$ and $\Lambda$ from the First Seven
  Supernovae at $z\geq 0.35$}},
  \href{https://doi.org/10.1086/304265}{\emph{Astrophys. J.} {\bfseries 483}
  (1997) 565}.

\bibitem{Perlmutter_1998}
S.~Perlmutter, G.~Aldering, M.D.~Valle, S.~Deustua et~al., \emph{{Discovery of
  a supernova explosion at half the age of the Universe}},
  \href{https://doi.org/10.1038/34124}{\emph{nature} {\bfseries 391} (1998)
  51}.

\bibitem{Perlmutter_1999}
S.~Perlmutter, G.~Aldering, G.~Goldhaber, R.A.~Knop et~al., \emph{{Measurements
  of $\Omega$ and $\Lambda$ from $42$ High-Redshift Supernovae}},
  \href{https://doi.org/10.1086/307221}{\emph{Astrophys. J.} {\bfseries 517}
  (1999) 565}.

\bibitem{Riess_2004}
A.G.~Riess, L.-G.~Strolger, J.~Tonry, S.~Casertano et~al., \emph{{Type Ia
  Supernova Discoveries at $z > 1$ from the Hubble Space Telescope: Evidence
  for Past Deceleration and Constraints on Dark Energy Evolution*}},
  \href{https://doi.org/10.1086/383612}{\emph{Astrophys. J.} {\bfseries 607}
  (2004) 665}.

\bibitem{Riess_2007}
A.G.~Riess, L.-G.~Strolger, S.~Casertano, H.C.~Ferguson et~al., \emph{{New
  Hubble Space Telescope Discoveries of Type Ia Supernovae at $z\geq 1$:
  Narrowing Constraints on the Early Behavior of Dark Energy*}},
  \href{https://doi.org/10.1086/510378}{\emph{Astrophys. J.} {\bfseries 659}
  (2007) 98}.

\bibitem{Hawkins_2003}
E.~Hawkins, S.~Maddox, S.~Cole, O.~Lahav et~al., \emph{{The 2dF Galaxy Redshift
  Survey: correlation functions, peculiar velocities and the matter density of
  the Universe}},
  \href{https://doi.org/10.1046/j.1365-2966.2003.07063.x}{\emph{Mon. Notices
  Royal Astron. Soc.} {\bfseries 346} (2003) 78}.

\bibitem{Max_2004}
M.~Tegmark, M.A.~Strauss, M.R.~Blanton, K.~Abazajian et~al., \emph{Cosmological
  parameters from sdss and wmap},
  \href{https://doi.org/10.1103/PhysRevD.69.103501}{\emph{Phys. Rev. D}
  {\bfseries 69} (2004) 103501}.

\bibitem{Cole_2005}
S.~Cole, W.J.~Percival, J.A.~Peacock, P.~Norberg et~al., \emph{{The 2dF Galaxy
  Redshift Survey: power-spectrum analysis of the final data set and
  cosmological implications}},
  \href{https://doi.org/10.1111/j.1365-2966.2005.09318.x}{\emph{Mon. Notices
  Royal Astron. Soc.} {\bfseries 362} (2005) 505}.

\bibitem{Eisenstein_2005}
D.J.~Eisenstein, I.~Zehavi, D.W.~Hogg, R.~Scoccimarro et~al., \emph{{Detection
  of the Baryon Acoustic Peak in the Large-Scale Correlation Function of SDSS
  Luminous Red Galaxies}},
  \href{https://doi.org/10.1086/466512}{\emph{Astrophys. J.} {\bfseries 633}
  (2005) 560}.

\bibitem{Nojiri_2007}
S.~Nojiri and S.D.~Odintsov, \emph{{Introduction to modified gravity and
  gravitational alternative for dark energy}},
  \href{https://doi.org/10.1142/S0219887807001928}{\emph{Int. J. Geom. Methods
  Mod. Phys.} {\bfseries 04} (2007) 115}.

\bibitem{Capozziello_2011}
S.~Capozziello and M.~{De Laurentis}, \emph{Extended theories of gravity},
  \href{https://doi.org/10.1016/j.physrep.2011.09.003}{\emph{Phys. Rep.}
  {\bfseries 509} (2011) 167}.

\bibitem{Saridakis_2021}
E.N.~Saridakis, R.~Lazkoz, V.~Salzano, P.V.~Moniz et~al., \emph{Modified
  Gravity and Cosmology: An Update by the CANTATA Network}, Springer Cham
  (2021),
  \href{https://doi.org/10.1007/978-3-030-83715-0}{10.1007/978-3-030-83715-0}.

\bibitem{Stelle_1977}
K.S.~Stelle, \emph{{Renormalization of higher-derivative quantum gravity}},
  \href{https://doi.org/10.1103/PhysRevD.16.953}{\emph{Phys. Rev. D} {\bfseries
  16} (1977) 953}.

\bibitem{Riess_1998}
A.G.~Riess, A.V.~Filippenko, P.~Challis, A.~Clocchiatti et~al.,
  \emph{{Observational Evidence from Supernovae for an Accelerating Universe
  and a Cosmological Constant}},
  \href{https://doi.org/10.1086/300499}{\emph{Astron. J.} {\bfseries 116}
  (1998) 1009}.

\bibitem{Betoule_2014}
M.~Betoule, R.~Kessler, J.~Guy, J.~Mosher et~al., \emph{{Improved cosmological
  constraints from a joint analysis of the SDSS-II and SNLS supernova
  samples}}, \href{https://doi.org/10.1051/0004-6361/201423413}{\emph{Astron.
  Astrophys.} {\bfseries 568} (2014) A22}.

\bibitem{Ade_2015}
P.A.R.~Ade, N.~Aghanim, M.~Arnaud, M.~Ashdown et~al., \emph{{Planck 2015
  results - XIII. Cosmological parameters}},
  \href{https://doi.org/10.1051/0004-6361/201525830}{\emph{Astron. Astrophys.}
  {\bfseries 594} (2015) A13}.

\bibitem{Aghanim_2016}
N.~Aghanim, M.~Arnaud, M.~Ashdown, J.~Aumont et~al., \emph{{Planck 2015 results
  - XI. CMB power spectra, likelihoods, and robustness of parameters}},
  \href{https://doi.org/10.1051/0004-6361/201526926}{\emph{Astron. Astrophys.}
  {\bfseries 594} (2016) A11}.

\bibitem{Carroll_2004}
S.M.~Carroll, V.~Duvvuri, M.~Trodden and M.S.~Turner, \emph{{Is cosmic speed-up
  due to new gravitational physics?}},
  \href{https://doi.org/10.1103/PhysRevD.70.043528}{\emph{Phys. Rev. D}
  {\bfseries 70} (2004) 043528}.

\bibitem{Hu_2007}
W.~Hu and I.~Sawicki, \emph{{Models of $f(R)$ cosmic acceleration that evade
  solar system tests}},
  \href{https://doi.org/10.1103/PhysRevD.76.064004}{\emph{Phys. Rev. D}
  {\bfseries 76} (2007) 064004}.

\bibitem{Sawicki_2007}
I.~Sawicki and W.~Hu, \emph{{Stability of cosmological solutions in $f(R)$
  models of gravity}},
  \href{https://doi.org/10.1103/PhysRevD.75.127502}{\emph{Phys. Rev. D}
  {\bfseries 75} (2007) 127502}.

\bibitem{Amendola_2008}
L.~Amendola and S.~Tsujikawa, \emph{{Phantom crossing, equation-of-state
  singularities, and local gravity constraints in $f(R)$ models}},
  \href{https://doi.org/10.1016/j.physletb.2007.12.041}{\emph{Phys. Lett. B}
  {\bfseries 660} (2008) 125}.

\bibitem{Capozziello_2008a}
S.~Capozziello, P.~Martin-Moruno and C.~Rubano, \emph{{Dark energy and dust
  matter phases from an exact $f(R)$-cosmology model}},
  \href{https://doi.org/10.1016/j.physletb.2008.04.061}{\emph{Phys. Lett. B}
  {\bfseries 664} (2008) 12}.

\bibitem{Nojiri_2007a}
S.~Nojiri and S.D.~Odintsov, \emph{{Introduction to Modified Gravity and
  Gravitational Alternative for Dark Energy}},
  \href{https://doi.org/10.1142/S0219887807001928}{\emph{Int. J. Geom. Methods
  Mod. Phys.} {\bfseries 4} (2007) 115}.

\bibitem{Multamaki_2004}
T.~Multam\"aki and I.~Vilja, \emph{{Spherically symmetric solutions of modified
  field equations in $f(R)$ theories of gravity}},
  \href{https://doi.org/10.1103/PhysRevD.74.064022}{\emph{Phys. Rev. D}
  {\bfseries 74} (2006) 064022}.

\bibitem{Multamaki_2007}
T.~Multam\"aki and I.~Vilja, \emph{{Static spherically symmetric perfect fluid
  solutions in $f(R)$ theories of gravity}},
  \href{https://doi.org/10.1103/PhysRevD.76.064021}{\emph{Phys. Rev. D}
  {\bfseries 76} (2007) 064021}.

\bibitem{Shamir_2010}
M.F.~Shamir, \emph{{Some Bianchi type cosmological models in $f(R)$ gravity}},
  \href{https://doi.org/10.1007/s10509-010-0371-5}{\emph{Astrophys. and Space
  Sci.} {\bfseries 330} (2010) 183}.

\bibitem{Santos_2007}
{Santos, J. and Alcaniz, J. S. and Rebou\ifmmode \mbox{\c{c}}\else
  \c{c}\fi{}as, M. J. and Carvalho, F. C.}, \emph{{Energy conditions in $f(R)$
  gravity}}, \href{https://doi.org/10.1103/PhysRevD.76.083513}{\emph{Phys. Rev.
  D} {\bfseries 76} (2007) 083513}.

\bibitem{Bohmer_2007}
O.~Bertolami, C.G.~B\"ohmer, T.~Harko and F.S.N.~Lobo, \emph{{Extra force in
  $f(R)$ modified theories of gravity}},
  \href{https://doi.org/10.1103/PhysRevD.75.104016}{\emph{Phys. Rev. D}
  {\bfseries 75} (2007) 104016}.

\bibitem{Harko_2008}
T.~Harko, \emph{{Modified gravity with arbitrary coupling between matter and
  geometry}},
  \href{https://doi.org/https://doi.org/10.1016/j.physletb.2008.10.007}{\emph{Phys.
  Lett. B} {\bfseries 669} (2008) 376}.

\bibitem{Harko_2010b}
T.~Harko and F.S.~Lobo, \emph{{$f(R, L_{m})$ gravity}},
  \href{https://doi.org/10.1140/epjc/s10052-010-1467-3}{\emph{Eur. Phys. J. C}
  {\bfseries 70} (2010) 373}.

\bibitem{Harko_2010}
T.~Harko, \emph{{Galactic rotation curves in modified gravity with nonminimal
  coupling between matter and geometry}},
  \href{https://doi.org/10.1103/PhysRevD.81.084050}{\emph{Phys. Rev. D}
  {\bfseries 81} (2010) 084050}.

\bibitem{Harko_2010a}
T.~Harko, \emph{{The matter Lagrangian and the energy-momentum tensor in
  modified gravity with nonminimal coupling between matter and geometry}},
  \href{https://doi.org/10.1103/PhysRevD.81.044021}{\emph{Phys. Rev. D}
  {\bfseries 81} (2010) 044021}.

\bibitem{Savvas_2009}
S.~Nesseris, \emph{{Matter density perturbations in modified gravity models
  with arbitrary coupling between matter and geometry}},
  \href{https://doi.org/10.1103/PhysRevD.79.044015}{\emph{Phys. Rev. D}
  {\bfseries 79} (2009) 044015}.

\bibitem{Valerio_2007}
V.~Faraoni, \emph{{Viability criterion for modified gravity with an extra
  force}}, \href{https://doi.org/10.1103/PhysRevD.76.127501}{\emph{Phys. Rev.
  D} {\bfseries 76} (2007) 127501}.

\bibitem{Valerio_2009}
V.~Faraoni, \emph{{Lagrangian description of perfect fluids and modified
  gravity with an extra force}},
  \href{https://doi.org/10.1103/PhysRevD.80.124040}{\emph{Phys. Rev. D}
  {\bfseries 80} (2009) 124040}.

\bibitem{Gonccalves_2023}
B.~Gon{\c{c}}alves, P.~Moraes and B.~Mishra, \emph{{Cosmology from Non-Minimal
  Geometry-Matter Coupling}},
  \href{https://doi.org/10.1002/prop.202200153}{\emph{Fortschritte der Phys.}
  {\bfseries 71} (2023) 2200153}.

\bibitem{Starobinsky_1980}
A.~Starobinsky, \emph{{A New Type of Isotropic Cosmological Models Without
  Singularity}},
  \href{https://doi.org/10.1016/0370-2693(80)90670-X}{\emph{Phys. Lett. B}
  {\bfseries 91} (1980) 99}.

\bibitem{Harko_2014}
T.~Harko and F.~Lobo, \emph{{Generalized Curvature-Matter Couplings in Modified
  Gravity}}, \href{https://doi.org/10.3390/galaxies2030410}{\emph{Galaxies}
  {\bfseries 2} (2014) 410}.

\bibitem{Jimenez_2002}
R.~Jimenez and A.~Loeb, \emph{{Constraining Cosmological Parameters Based on
  Relative Galaxy Ages}},
  \href{https://doi.org/10.1086/340549}{\emph{Astrophys. J.} {\bfseries 573}
  (2002) 37}.

\bibitem{Narawade_2023}
S.A.~Narawade and B.~Mishra, \emph{{Phantom Cosmological Model with
  Observational Constraints in $f(Q)$ Gravity}},
  \href{https://doi.org/10.1002/andp.202200626}{\emph{Annalen Phys.} (2023)
  2200626}.

\bibitem{Brout_2022}
D.~Brout, D.~Scolnic, B.~Popovic, A.G.~Riess et~al., \emph{{The $Pantheon^{+}$
  Analysis:Cosmological Constraints}},
  \href{https://doi.org/10.3847/1538-4357/ac8e04}{\emph{Astrophys. J.}
  {\bfseries 938} (2022) 110}.

\bibitem{Aviles_2012}
A.~Aviles, C.~Gruber, O.~Luongo and H.~Quevedo, \emph{{Cosmography and
  constraints on the equation of state of the Universe in various
  parametrizations}},
  \href{https://doi.org/10.1103/PhysRevD.86.123516}{\emph{Phys. Rev. D}
  {\bfseries 86} (2012) 123516}.

\bibitem{Gruber_2014}
C.~Gruber and O.~Luongo, \emph{{Cosmographic analysis of the equation of state
  of the universe through Pad\'e approximations}},
  \href{https://doi.org/10.1103/PhysRevD.89.103506}{\emph{Phys. Rev. D}
  {\bfseries 89} (2014) 103506}.

\bibitem{Weinberg_1971}
S.~Weinberg and R.V.~Wagoner, \emph{{Gravitation and Cosmology: Principles and
  Applications of the General Theory of Relativity}}, {\emph{Phys. Today}
  {\bfseries 26} (1973) 57}.

\bibitem{Visser_2004}
M.~Visser, \emph{{Jerk, snap and the cosmological equation of state}},
  \href{https://doi.org/10.1088/0264-9381/21/11/006}{\emph{Class. Quant. Grav.}
  {\bfseries 21} (2004) 2603}.

\bibitem{Visser_2005}
M.~Visser, \emph{{Cosmography: Cosmology without the Einstein equations}},
  \href{https://doi.org/10.1007/s10714-005-0134-8}{\emph{Gen. Relativ. Gravit.}
  {\bfseries 37} (2005) 1541}.

\bibitem{Yang_2020}
Y.~Yang and Y.~Gong, \emph{{The evidence of cosmic acceleration and
  observational constraints}},
  \href{https://doi.org/10.1088/1475-7516/2020/06/059}{\emph{JCAP} {\bfseries
  2020} (2020) 059}.

\bibitem{Capozziello_2015}
S.~Capozziello, O.~Luongo and E.N.~Saridakis, \emph{{Transition redshift in
  $f(T)$ cosmology and observational constraints}},
  \href{https://doi.org/10.1103/PhysRevD.91.124037}{\emph{Phys. Rev. D}
  {\bfseries 91} (2015) 124037}.

\bibitem{Sahni_2003}
V.~Sahni, T.D.~Saini, A.A.~Starobinsky and U.~Alam, \emph{{Statefinder-A new
  geometrical diagnostic of dark energy}},
  \href{https://doi.org/10.1134/1.1574831}{\emph{J. Exp. Theor. Phys.}
  {\bfseries 77} (2003) 201}.

\bibitem{Alam_2003}
U.~Alam, V.~Sahni, T.~Deep~Saini and A.A.~Starobinsky, \emph{{Exploring the
  expanding Universe and dark energy using the statefinder diagnostic}},
  \href{https://doi.org/10.1046/j.1365-8711.2003.06871.x}{\emph{Mon. Notices
  Royal Astron. Soc.} {\bfseries 344} (2003) 1057}.

\bibitem{Ding_2015}
X.~Ding, M.~Biesiada, S.~Cao, Z.~Li and Z.-H.~Zhu, \emph{{Is there evidence for
  dark energy evolution?}},
  \href{https://doi.org/10.1088/2041-8205/803/2/L22}{\emph{Astrophys. J. Lett.}
  {\bfseries 803} (2015) L22}.

\bibitem{Zheng_2016}
X.~Zheng, X.~Ding, M.~Biesiada, S.~Cao and Z.-H.~Zhu, \emph{{What are the
  $Omh^{2}(z_{1},z_{2})$ and $Om(z_{1},z_{2})$ diagnostics telling us in light
  of $H(z)$ data?}},
  \href{https://doi.org/10.3847/0004-637X/825/1/17}{\emph{Astrophys. J.}
  {\bfseries 825} (2016) 17}.

\bibitem{Qi_2018}
J.-Z.~Qi, S.~Cao, M.~Biesiada, T.-P.~Xu, Y.~Wu, S.-X.~Zhang et~al., \emph{{What
  do parameterized $Om(z)$ diagnostics tell us in light of recent
  observations?}}, \href{https://doi.org/10.1088/1674-4527/18/6/66}{\emph{Res.
  Astron. Astrophys.} {\bfseries 18} (2018) 066}.

\bibitem{VAGNOZZI202227}
S.~Vagnozzi, F.~Pacucci and A.~Loeb, \emph{{Implications for the Hubble tension
  from the ages of the oldest astrophysical objects}},
  \href{https://doi.org/https://doi.org/10.1016/j.jheap.2022.07.004}{\emph{J.
  High Energy Phys.} {\bfseries 36} (2022) 27}.

\end{thebibliography}\endgroup

\end{document}